\documentclass[twocolumn]{autart}    

\usepackage{graphicx}          
\usepackage{psfrag,dblfloatfix}
\usepackage{amsmath,amssymb,amsfonts,mathrsfs}
\usepackage{color,epstopdf}
\usepackage{enumerate} 
\usepackage{subfigure}
\usepackage{float}
\usepackage[authoryear,sort]{natbib}   
\usepackage[utf8]{inputenc} 
\usepackage[T1]{fontenc}
\usepackage{verbatim}
\usepackage{multicol,setspace}

\newcommand{\R}{\mathbb{R}}
\newcommand{\Z}{\mathbb{Z}}
\newcommand{\Zzero}{\mathbb{Z}_{\ge 0}}

\newcommand{\eps}{\varepsilon}
\def\qedp{\hspace*{\fill}~{\tiny $\blacksquare$}}

\definecolor{darkgreen}{rgb}{0.0, 0.55, 0.0}
\definecolor{amaranth}{rgb}{0.9, 0.17, 0.31}
\usepackage[prependcaption,colorinlistoftodos]{todonotes}

\newtheorem{theorem}{Theorem}
\newtheorem{itlemma}{Lemma}
\newtheorem{itdefinition}{Definition}
\newtheorem{itproposition}{Proposition}
\newtheorem{itresult}{Result}
\newtheorem{itremark}{Remark}
\newtheorem{itassumption}{Assumption}
\newtheorem{itcorollary}{Corollary}
\newtheorem{itexample}{Example}

\newenvironment{proposition}{\begin{itproposition}\rm}{\end{itproposition}}
\newenvironment{remark}{\begin{itremark}\rm}{\end{itremark}}

\newenvironment{lemma}{\begin{itlemma}\rm}{\end{itlemma}}
\newenvironment{corollary}{\begin{itcorollary}\rm}{\end{itcorollary}}

\allowdisplaybreaks

\begin{document}

\newpage
\begin{frontmatter}

\title{Quantized State Feedback Stabilization of Nonlinear Systems under Denial-of-Service\thanksref{footnoteinfo}} 

\thanks[footnoteinfo]{
	This work was supported in part by the ``RevealFlight" ARC at UCLouvain, and in part	
	by JST CREST Grant No.~JPMJCR15K3 and
	by JSPS under Grant-in-Aid for Scientific Research~(B) Grant
	No.~18H04020. The material in this paper is to be partially presented at
	the 24th International Symposium on Mathematical Theory of Networks and
	Systems (MTNS), Cambridge, UK, August 23-27, 2021.
Corresponding author: Shuai Feng. }

\author[First]{Mingming Shi}\ead{mingming.shi@uclouvain.be},    
\author[Second]{Shuai Feng} \ead{s.feng@rug.nl},               
\author[Third]{Hideaki Ishii}\ead{ishii@c.titech.ac.jp}  

\address[First]{UCLouvain, Louvain-la-Neuve, 1348, Belgium}  
\address[Second]{ENTEG, Faculty of Science and Engineering, University of Groningen, the Netherlands}                               
\address[Third]{Department of Computer Science, Tokyo Institute of Technology, Yokohama, Japan}             


\begin{abstract}                          
This paper studies the resilient control of networked systems in the presence of cyber attacks. In particular, we consider the state feedback stabilization problem for nonlinear systems when the state measurement is sent to the controller via a communication channel that only has a finite transmitting rate and is moreover subject to cyber attacks in the form of Denial-of-Service (DoS). We use a dynamic quantization method to update the quantization range of the encoder/decoder and characterize the number of bits for quantization needed to stabilize the system under a given level of DoS attacks in terms of duration and frequency. Our theoretical result shows that under DoS attacks, the required data bits to stabilize nonlinear systems by state feedback control are larger than those without DoS since the communication interruption induced by DoS  makes the quantization uncertainty expand more between two successful transmissions. Even so,  in the simulation, we show that the actual quantization bits can be much smaller than the theoretical value. 
\end{abstract}

\end{frontmatter}

\section{Introduction}
This paper focuses on the resilient control of Cyber-physical systems (CPSs) under cyber attacks initiated by adversarial attackers, which can greatly affect realtime data exchange between communication devices in CPSs. Problems of using general purpose channels such as the Internet and wireless communication for feedback control purposes have been extensively studied, for example in \citet{Antsaklis-specialissues,walsh2001asymptotic,Wang-PerETC}. However, the application of such communication channels also creates vulnerabilities in CPSs \citep{cheng2017guest}. Recently, various studies to analyze and mitigate their effects have been carried out; for overviews on this topic, see, for example, \citet{teixeira2015secure,Fawzi-Sesuresest,Mo-CPS-smrtgrid,Pasqualetti-Attackdetection,de2015input} and the references therein.
	

Among different classes of cyber attacks, in this paper, we focus on Denial-of-Service (DoS) attacks, which 
disrupt and block communication over channels temporarily. 
In particular, we consider the problem of stabilizing nonlinear systems by state feedback, where the measurement of the process state is transmitted through a communication channel with limited data rate and DoS attacks. This implies that the state measurement should first be quantized before transmission and some packets carrying the state information may not be received by the receiver under DoS attacks. For the stability of networked control systems, it is well recognized that there are fundamental limitations on the communication data rate \citep{liberzon2003hybrid,liberzon2005stabilization,LingbitrateEvent,nair2004topological,Tatikonda_commconst}. It should be mentioned that stabilization under stochastic packet dropping has been studied by assuming that the packet dropping follows certain probability distributions \citep{amin2009safe,Gupta_TacErasurechannels}. In \citet{Okano2014Finitedata,You2011mindata}, communication constraints on both data rate and such stochastic packet losses have been studied. However, as malicious attackers can schedule the DoS attacks deliberately, the obtained results would not be applicable when the packet drops are induced by DoS attacks. This poses new challenges in theoretical analysis and controller design.

For nonlinear systems, system stabilization under state quantization has been intensively studied in
	the literature (see, e.g., \citet{de2006nonlinear,depersis2004,nair2004topological}). In general, dynamic quantization methods are proposed to achieve asymptotic stabilization. It has been shown that quantized feedback stabilizability of nonlinear systems relies on properties of the closed-loop systems without state quantization, for example, input-to-state stability \citep{liberzon2003hybrid,liberzon2005stabilization}, integral input-to-state stabilizability \citep{de2006nonlinear}, or only state feedback stabilizablity \citep{depersis2004}. Moreover, the work \citet{nair2004topological} has shown that a nonlinear system is locally uniformly asymptotically stabilizable if and only if the data rate exceeds the plant's local topological feedback entropy at the equilibrium.


More recently, system stabilization under DoS attacks has drawn the attention of researchers. Specifically, in \citet{de2015input}, the authors have proposed a deterministic framework to model DoS attack signals by characterizing their frequency and duration. There, it is proved that the closed-loop system is stable if the accumulation of system's stable mode during DoS-free time outperforms the counterpart of unstable mode under DoS signals. Although following this framework, many problems on control under DoS have been investigated by, e.g., \citet{LuMutichannelDoS,feng2017,cetinkaya2019overview}, very little attention has been paid to nonlinear systems, especially considering the generality of nonlinear systems in the real applications \citep{de2016networked}. 

We emphasize that it is more challenging to stabilize systems considering both limited data rate and DoS attacks as the latter makes the prediction of the sampling time instants difficult, which clearly affects the characterization of the quantization uncertainty. A key question is how to select/how many should be the bits of the quantizer such that quantization error converges to zero eventually. There are several works which investigate stabilization of linear systems with quantized state feedback such as \citet{wakaiki2017quantized,LingbitrateEvent,Shuai_Stabilizationun}. They show that the data rate to stabilize the systems under DoS attacks should be no less than that without DoS amplified by a term related to the frequency and duration of DoS attacks. 

However, there are still not many comparable results for nonlinear systems. Even though some papers have studied state estimation over communication channels with finite capacity and packet erasure \citep{diwadkar2013limitations,sanjaroon2018estimation}, these results are all derived in stochastic scenarios, which as mentioned before, may not be suitable for networked systems under DoS attacks. The recent work of \citet{Kato2019} deals with nonlinear systems from an alternative viewpoint, exposing the limitation of control based on linearization
in the presence of DoS attacks..

In this paper, we study the deterministic stabilization of nonlinear systems with quantized state feedback and DoS attacks under the same framework as in \citet{de2015input}. Specifically, we assume that the communication network has a finite data rate. Hence the state measurement should be quantized before transmission. Due to the presence of DoS attacks, some transmissions may fail. We show that if the number of transmission bits is larger than a value which depends on the system Lipschitz constant and the frequency and duration of DoS attacks, then the system can be stabilized (asymptotically to the origin). Confirming with the intuition, the predicted quantization bits for nonlinear systems stabilization under DoS attacks are larger than those for the attack free case. Our result relies only on the assumption that the system is state feedback stabilizable as in \citet{depersis2004}. 

 The rest of the paper is organized as follows. In Section \ref{sec:problem}, we introduce the considered networked system and the DoS model, and provide the problem formulation. In Section \ref{sec:dynquant}, we design the encoder and decoder for the state transmission and provide the quantized state feedback controller. In Section \ref{subsec:asyest}, we show a result guaranteeing that the quantization error under DoS decreases gradually with a sufficient number of quantization bits. Section \ref{sec:asymtoticsta} proves asymptotic stabilization of the nonlinear system under DoS by the proposed quantized state feedback control. Section \ref{sec:examp} discusses a simulation example and Section \ref{sec:conclusion} concludes the paper. Compared with the preliminary conference version \citep{MingmingDoSns}, the current paper provides all the technical proofs, and more thorough discussions on the results.




\section{Problem formulation}\label{sec:problem}

\subsection{Notation}
We let $\R$ and $\R_{\ge 0}$ denote the sets of real numbers and reals no smaller than $0$, respectively. Given $a\in \R$, let $\Z$ and $\Z_{\ge a}$ denote the sets of integers, and integers no smaller than $a$, respectively. We let $|a|$ denote its absolute value. Given a vector $b=[b_1\ b_2\ \cdots\ b_c]^\top\in \R^c$ with $c\in \Z_{\ge 1}$, we let $|b|_\infty$ denote its infinity norm, namely $|b|_\infty=\max_{i}|b_i|$. Given $d\in \R_{>0}$, we let $\mathcal{B}(d)$ denote the hypercube that is centered at the origin and has length of $2d$ for each edge. For a vector $h\in \R^c$ and a real vector set $\mathcal{S}$, we let their sum be $h+\mathcal{S}=\{h+\xi:\xi\in \mathcal{S}\}$. 
For an exponentially decaying function $c\text{e}^{-\lambda t}$ where $t\in \R_{\ge 0}$ is the argument and $c\in \R$, $\lambda\in \R_{> 0}$ are constants, we say $\lambda$ is the decay rate of this function. A function $f:[0,\infty)\rightarrow [0,\infty)$ is said to be of class $\mathcal K$ if it is continuous, strictly increasing and $f(0)=0$. Furthermore, if $f(a)\rightarrow \infty$ when $a\rightarrow \infty$, then it is said to be of class $\mathcal K_\infty$. 
\subsection{Quantized Feedback Stabilization of Nonlinear Systems}\label{sec:QFSNS}

Consider the following nonlinear system 
\begin{equation}\label{eq:sys}
\dot x=f(x,u)
\end{equation}
where $x\in \R^n$ denotes the state and $u\in \R^m$ is the control input. Here, $f(\cdot,\cdot)$ is a smooth map and satisfies $f(0,0)=0$. For this system, we have the following assumption.
\begin{assum}\label{asm:stability}
	There exists a smooth feedback control law given as 
	\begin{equation}\label{eq:standcont}
	u=k(x)
	\end{equation}
	such that system \eqref{eq:sys} is globally asymptotically stable.
\end{assum}

As a consequence of Assumption \ref{asm:stability}, there exist a smooth Lyapunov function $V(\cdot):\R^n\rightarrow \R_{\ge 0}$ and class $\mathcal{K}_\infty$ functions $\alpha_1(\cdot)$, $\alpha_2(\cdot)$ and $\alpha(\cdot)$ such that
\begin{align}
&\alpha_1(|x|_\infty)\le V(x)\le \alpha_2(|x|_\infty)\nonumber\\
&\frac{\partial{V}}{\partial{x}}f(x,k(x))\le -\alpha(|x|_\infty). \nonumber
\end{align}

In this system, we assume that the plant and the actuator are co-located while the sensors are remote from the plant controller. The sensors can measure the exact plant state. After each sampling, the sensors send the system state to the controller via a communication channel, which can transmit data with only finite rate. This implies that there should be a quantization mechanism for transmitting the measured state. Hence we assume that the sensor/controller is embedded with the encoder/decoder. 

Moreover, the state is sampled and transmitted in discrete time. We let $\{t_i\}_{i\in \Zzero}$ be the sequence of time instants at which the sensors measure, encode and send the state of the plant. During the encoding-decoding process, we assume that the state is sampled in a periodic manner. Hence there exists a constant sampling period $\Delta$ such that $t_i=i\Delta$ for $i\in \Zzero$. 

After receiving the quantized state information, the decoder updates its state estimate. We assume that the sensors can finish the state measuring and encoding immediately; and the actuator can decode and apply the control input without delay. We also assume that the communication protocol is acknowledgment (ACK)-based and there are no delays in ACK and state transmissions.

We use dynamic quantization methods for the state transmission, which contains two stages generally. In the \emph{zooming out} stage, the quantization range is enlarged to capture the system state. On the other hand, in the \emph{zooming in} stage, the quantization range and the state estimate error decrease.
By the evolution of zooming-in, the estimation error converges to zero asymptotically, and therefore the system is stabilized asymptotically. In this paper, we adopt the following assumption that the encoder and decoder have a common knowledge of the initial state.
\begin{assum}\label{asm:statebd}
	The initial state satisfies $|x_0|_\infty\le X$ where $X$ is known by the encoder and the decoder.
\end{assum}

\subsection{Time-constrained Denial-of-Service}

In general, communication channels do not only suffer from transmission rate constraints, but also are affected by other problems like noises or packet losses. In this paper, we focus on the case in which the packet drops are induced by DoS. DoS may be caused by legitimate but mass communication or by intentional adversary attacks. In this paper, we do not distinguish them and consider DoS as malicious DoS attacks launched by adversary attackers. In the following, we introduce the DoS model that was first proposed in \citet{de2015input}.

Let $\{h_k\}_{k\in \Zzero}$ with $h_0\ge t_0$ represent the sequence of time instants when the network changes from nominal status to DoS status. For each transition $h_k$, let $\tau_k\ge 0$ be the length of the DoS interval. Then the $k\rm {th}$ DoS time interval can be represented as
\begin{equation*}
H_k=\{h_k\}\cup[h_k,h_k+\tau_k[.
\end{equation*}
If $\tau_k=0$, then $H_k$ is a pulse. Given $\tau,t$ with $\tau\le t$, let $k(\tau,t)$ denote the number of DoS transitions from absence to presence over $[\tau,t]$. Thus,  
\begin{equation*}
\Xi(\tau,t)=\bigcup\limits_{k\in \Zzero} H_k\cap [\tau,t]
\end{equation*}
denotes the subset of $[\tau,t]$ where DoS is on. We adopt the following assumptions to characterize the DoS frequency and duration, which are able to capture several different packet dropping situations \citep{de2015input}.
\begin{assum}[DoS Frequency]\label{asp:DoSFreq}
	There exist constants $\eta\in \R_{\ge 0}$ and $\tau_D\in \R_{>0}$ such that 
	\begin{equation*}\label{eq:DoSF}
	k(\tau,t)\le \eta+\frac{t-\tau}{\tau_D}
	\end{equation*}
	for all $\tau,t\ge 0$ with $t\ge \tau$.
\end{assum}
\begin{assum}[DoS Duration]\label{asp:DoSDurat}
	There exist constants $\kappa\in \R_{\ge 0}$ and $T\in \R_{>0}$ such that 
	\begin{equation*}\label{eq:DoSD}
	\Xi(\tau,t)\le \kappa+\frac{t-\tau}{T}
	\end{equation*}
	for all $\tau,t\ge 0$ with $t\ge \tau$.
\end{assum}

Due to the presence of DoS, the state may not be successfully received by the controller at the nominal transmission times. Let $\{z_{\ell}\}_{\ell\in \Zzero}$ be the sequence of successful transmission instants. The following result characterizes the relation between the successful transmission time instants and the number of successful transmissions.
\begin{lemma}\citep{Shuai_Stabilizationun}\label{lem:zmbd}
	For periodic transmissions with period $\Delta$ and DoS sequences satisfying Assumptions \ref{asp:DoSFreq} and \ref{asp:DoSDurat}, if $
	\sigma :=1-\frac{1}{T}-\frac{\Delta}{\tau_D}>0 $, then
	\begin{align}
	z_0&\le (\kappa+\eta \Delta)/\sigma \label{eq:z0bound} \\ 
	z_\ell-z_0&\le  (\ell \Delta+\kappa+\eta \Delta)/\sigma  \label{eq:zmbound}.
	\end{align}
	\qedp
\end{lemma}

\section{Quantized State Feedback}\label{sec:dynquant}
In this section, we detail the state quantization algorithm and the feedback control. We equip the encoder/decoder with two variables: the state estimate $\overline x(t)\in \R^n$ and the quantization range $L(t)\in \R_{\ge 0}$ that upper bounds the infinite norm of the estimation error $e_x(t)=\overline x(t)-x(t)$. To facilitate the presentation of the encoded state feedback, we first introduce two positive real numbers $W$ and $O>W$, characterizing the evolution regions of the system state $x$ and the state estimate $\overline x\in \R^n$, respectively. Also let $U>0$ be the bound on the control input. Overall, one has 
\begin{equation}\label{eq:xubound}
|x|_\infty\le W,\quad |\overline{x}|_\infty\le O, \quad |u|_\infty\le U.
\end{equation}
Later we will prove that the system never leaves these regions under the proposed control scheme.
Then let $F$ be the Lipschitz constant such that 
\begin{equation}\label{eq:FOineq}
|f(x,u)-f(\overline x,u)|_\infty\le F|x-\overline x|_\infty
\end{equation}
is valid for all $x$, $\overline x$ and $u$ satisfying condition \eqref{eq:xubound}. Note by definition, $F$ depends on $W$, $O$ and $U$. How to select the values of $W,O$ and $U$ is postponed to Section \ref{sec:asymtoticsta}.

The state estimate and quantization range define the quantization region $\mathcal{S}(t)$ given by \citet{depersis2004}
\begin{equation}
\mathcal{S}(t)=\overline x(t)+\mathcal{B}\left(\frac{L(t)}{2}\right)
\end{equation}
which must contain the actual state $x(t)$. Hence the state is located within the quantization region when the estimation error satisfies
\begin{equation}\label{eq:qbound}
|e_x(t)|_\infty\le \frac{L(t)}{2}.
\end{equation}


Based on the state estimate, the feedback control applied to the plant is given by
\begin{equation}\label{eq:control}
u(t)=k(\overline x(t)).
\end{equation}
We let the initial condition of $\overline x$ be $\overline x(t_0^-)=0$ and the state estimate between two successful state transmissions evolve as follows
\begin{equation}\label{eq:dynxest}
\dot {\overline{x}}(t)=f(\overline x(t),k(\overline x(t))), \quad t\in [z_l, z_{l+1}[.
\end{equation}

As the system evolves, the state estimate uncertainty may increase. We need to update $\mathcal{S}(t)$ such that the state is always inside it since otherwise overflow will occur. 
We classify the updates of $\mathcal{S}(t)$ into two cases based on whether the time $t$ is before or after the instant of the first successful transmission $z_0$.

\emph{Case a)} $t<z_0$. Before $z_0$, all the transmission attempts fail. In view of \eqref{eq:dynxest}, $\overline x(t_0^-)=0$ implies
\begin{equation*}
\overline x(t)=0,\quad u(t)=0, \quad \forall t\in[t_0,z_0[.
\end{equation*}

Let $L(t_0^-)=2X $ be the initial value of the quantization range. For $t\in[t_0,z_0[$, the quantization range is updated as
\begin{equation}\label{eq:L<z0}
L(t)=2\phi_{\max}(t)
\end{equation}
where 
\begin{equation}\label{eq:phit}
\phi_{\max}(t)=\max\limits_{|\tilde x(t_0)|_\infty\le X, t'\in [t_0, t]}|\phi(\tilde x({t_0}),t',t_0)|_\infty
\end{equation}
with  $\phi(\tilde x(t_0),t',t_0)$ denoting the solution of system \eqref{eq:sys} at time $t'$ under initial state condition $\tilde x(t_0)$ and zero control input. Note that $\phi_{\max}(t)$ is accessible to both the encoder and decoder given the range of the initial state $X$ and the system dynamics, and hence it is possible for the encoder/decoder to update $L(t)$ as \eqref{eq:L<z0}.

\emph{Case b)} $t\ge z_0$. At $z_{\ell}$ with $ \ell\in \Zzero$, the quantization region $\mathcal{S}(z_{\ell})$ is characterized by $\overline x(z_{\ell})$ and $L(z_{\ell})$. For $t\in [z_\ell, z_{\ell+1}[$, the quantization range evolves according to the continuous-time dynamics
\begin{equation}\label{eq:Ldyn}
\dot L(t)=FL(t).
\end{equation}

Assume that at $t=z_{\ell+1}^-$, the system state is still within the quantization region. The sensor converts the state information to $nR$ bits by partitioning each edge of the quantization region $\mathcal{S}(z_{\ell+1}^-)$ into $2^R$ segments. This partition yields $2^{nR}$ sub-hypercubes and each of them can be represented by an $nR$ bits digital number. By sending a specific digital number, the decoder is informed about the sub-hypercube in which the state $x(z_{\ell+1}^-)$ lies. The encoder and decoder then take the centroid of this sub-hypercube as $\overline x(z_{\ell+1})$. 
Meanwhile, the quantization range at $t=z_{\ell+1}$ is set as 
\begin{equation}\label{eq:L_reset}
L(z_{\ell+1})=\frac{L(z_{\ell+1}^-)}{2^{R}}.
\end{equation}
This update implies that after a successful state transmission at $z_{\ell+1}$, the estimate error satisfies
\begin{equation}\label{eq:qerr_zk}
|e_x(z_{\ell+1})|_\infty\le \frac{L(z_{\ell+1}^-)}{2^{R+1}}=\frac{L(z_{\ell+1})}{2}.
\end{equation}  
%


\section{Asymptotic Estimation with DoS}\label{subsec:asyest}

As mentioned before, the estimate uncertainty may enlarge between two successful transmissions. When there are DoS attacks in the network, the communication becomes aperiodic. Hence, the estimate uncertainty expands unpredictably and may enlarge more than that of the nominal transmission. To compensate the additional expansion of the estimate uncertainty, we need to partition the quantization region $\mathcal{S}(t)$ into a larger number of sub-hypercubes. 

This intuition is formalized as the key result below. It shows that although the communication is affected by DoS, if the number of bits used in the quantization is sufficiently large, then the state can be estimated by the decoder asymptotically, under the assumption of the boundedness of state and control input (5).
\begin{proposition}\label{lem:qrangebitcon}
	Given any $X\in \R_{> 0}$, let $\overline z_0=(\kappa+\eta \Delta)/\sigma\ge z_0$ and choose $W$ and $O$ as
	\begin{equation}\label{eq:WO}
	W> \phi_{\max}(\overline z_0), \quad
	O=\alpha_1^{-1}\circ \alpha_2\left(2 W\right)
	\end{equation}
	respectively. 
	Suppose that the solution of system \eqref{eq:sys} with initial condition $|x(0)|_\infty\le X$ and control input $u(\cdot)$ for which $|u(t)|_\infty\le U$ for all $t\ge 0$ satisfies $|x(t)|\le W$ for all $t\ge 0$. 
	If the number of the quantization bits $B=nR $ is chosen such that
	\begin{equation}\label{eq:lem2con}
	R>\max\left\{\frac{F\Delta}{\sigma\ln (2)},\frac{F(\kappa+\eta\Delta)}{\sigma\ln (2)}\right\},
	\end{equation}
	then the estimate $\overline x(\cdot)$ generated by the decoder exists for all $t\ge 0$ and satisfies $|\overline x(t)|_\infty\le O$. Moreover, the state estimate error satisfies
	the inequalities
	\begin{align}
	|e_x(t)|_\infty
	&\le \phi_{\max}(t)<W,\ \ \forall t\in [{t_0},z_0[  \label{eq:lemqerr<z0} \\
	|e_x(t)|_\infty &\le \gamma\lambda^{\ell+1},\ \ \forall t\in [z_{\ell},z_{\ell+1}[ \label{eq:lemqerr>z0}
	\end{align}
	with 
	\begin{align}\label{eq:lem_gamlam}
	\gamma &= W\text{e}^{\frac{F(\kappa+\eta \Delta)}{\sigma}}\nonumber\\
	\lambda &= \text{e}^{(F\Delta/\sigma-R\ln 2)}<1.
	\end{align}
\end{proposition}
\noindent \emph{Proof.} First, we make the following two claims: \\
\emph{Claim a)} $|\overline x(t)|_\infty \le O$ for all $t\in [t_0, z_0[$.\\
\emph{Claim b)} For all $\ell\in \Zzero$, it holds $|\overline x(z_{\ell})|_
\infty\le 2W\le O$, $|\overline x(t)|_\infty\le O$ for all $t\in ]z_{\ell},z_{\ell+1}[$, and
\begin{equation}\label{eq:Lzl}
L(z_{\ell})=\frac{\text{e}^{F(z_{\ell}-z_0)}}{2^{R(\ell+1)}}L(z_{0}^-).
\end{equation}

Since $\overline x(t)=0$ for $t\in [t_0,z_0[$, it is clear that
\begin{align}\label{eq:errt<z0}
|e(t)|_\infty &= |x(t)-\overline x(t)|_\infty=|x(t)|_\infty\nonumber\\
& \le  \phi_{\max}(t)\le \phi_{\max}(z_0^-)< W
\end{align}
which proves the inequality \eqref{eq:lemqerr<z0} and hence Claim a).

For the second claim, we prove it by induction. We first show that $\overline x(z_0)$ exists and satisfies $|\overline x(z_0)|\le 2W$. In fact, by \eqref{eq:errt<z0} and \eqref{eq:L<z0}, the estimate error satisfies
$|e(t)|_\infty\le \frac{L(t)}{2}$ for all $ t\in [t_0,z_0[$. Hence $x(z_{0}^-)$ lies within the quantization region $\mathcal{S}(z_{0}^-)$. Then according to \eqref{eq:L_reset} and \eqref{eq:qerr_zk}, after the state is successfully transmitted for the first time, the quantization range and the estimate error satisfy
\begin{equation}
\begin{split}
L(z_0)=&\frac{L(z^-_0)}{2^R}\\
|e_x(z_0)|_\infty\le& \frac{L(z_0^-)}{2^{R+1}}=\frac{L(z_0)}{2}. \label{eq:errLz0}
\end{split}
\end{equation}
Moreover, the state estimate at $z_0$ satisfies
\begin{align}\label{eq:ovexz0}
|\overline x(z_0)|_\infty &\le |x(z_0)|_\infty+|e_x(z_0)|_\infty \nonumber\\
&\le  \phi_{\max}(z_0)+\frac{L(z_0^-)}{2^{R+1}}\nonumber\\
&\le\left(1+\frac{1}{2^{R}}\right)\phi_{\max}(z_0^-) \nonumber\\
&\le  \left(1+\frac{1}{2^{R}}\right)W<2W 
\end{align}
where the second inequality comes from \eqref{eq:phit} and \eqref{eq:errLz0}, the third from \eqref{eq:L<z0}, the fourth from \eqref{eq:WO}, and the last one from the fact that $\frac{1}{2^{R}}\le 1$. This shows that $|\overline x(z_0)|_\infty\le 2 W$.

Suppose now that the state estimate at $z_{\ell}$ with $\ell\in \Z_{\ge 0}$ exists and satisfies $ |\overline x(z_\ell)|_\infty\le 2 W<O$, and suppose that the quantization range satisfies \eqref{eq:Lzl}. In view of the fact that the state is successfully transmitted at $z_{\ell}$, we have
\begin{equation}\label{eq:err_zl}
|e_x(z_{\ell})|_\infty\le \frac{L(z_{\ell-1}^-)}{2^{R+1}}=\frac{L(z_{\ell})}{2}.
\end{equation}
Consider the evolution of $\overline x(t)$ for $[z_\ell,z_{\ell+1}[$, Assumption
1 implies that system \eqref{eq:dynxest} satisfies
\begin{equation*}
\dot V(\overline x(t))\le -\alpha(\overline x(t))<0,\ t\in [z_\ell,z_{\ell+1}[.
\end{equation*}
Hence for $t\in [z_\ell,z_{\ell+1}[$, we have
\begin{align}\label{eq:xtlbound}
|\overline x(t)|_\infty &\le \alpha_1^{-1}\circ V(\overline x(t)) <\alpha_1^{-1}\circ V(\overline x(z_{\ell}))\nonumber\\
&< \alpha_1^{-1}\circ \alpha_2 (|\overline x(z_{\ell})|_\infty)<\alpha_1^{-1}\circ \alpha_2 (2W)\nonumber\\
&= O
\end{align}
where the last equality follows from \eqref{eq:WO}. The boundedness of $x(t)$, $u(t)$ and \eqref{eq:xtlbound} implies that the Lipschitz constant $F$ in \eqref{eq:FOineq} remains to hold for $t\in [z_{\ell},z_{\ell+1}[$.  Hence the evolution of the norm of the estimate error can be bounded as
\begin{equation}
\frac{d|e_x(t)|_\infty}{dt} \le   |f(x,u)-f(\overline x,u)|_\infty \le   F |e_x(t)|_\infty
\end{equation}
where the last inequality follows from \eqref{eq:FOineq}. This implies
\begin{equation}\label{eq:qerr_xt}
|e_x(t)|_\infty\le \text{e}^{F(t-z_0)}|e_x(z_{\ell})|_\infty
\end{equation}
for all $t\in [z_{\ell},z_{\ell+1}[$. By \eqref{eq:Ldyn}, the quantization range satisfies
\begin{equation}\label{eq:Ltzl}
L(t)=\text{e}^{F(t-z_{\ell})}L(z_{\ell}), \quad \forall t\in [z_{\ell},z_{\ell+1}[.
\end{equation}
Combining \eqref{eq:err_zl}, \eqref{eq:qerr_xt} and \eqref{eq:Ltzl}, we have
\begin{equation*}\label{eq:errt}
|e_x(t)|_\infty\le \frac{L(t)}{2},\quad \forall t\in [z_{\ell},z_{\ell+1}[.
\end{equation*}
This holds true also at $t=z_{\ell+1}^-$, and hence the state lies within the quantization region $\mathcal{S}(t)$ at $t=z_{\ell+1}^-$. Then according to \eqref{eq:L_reset}, \eqref{eq:err_zl} and \eqref{eq:qerr_zk}, after the successful state transmission at $z_{\ell+1}$, the quantization range satisfies
\begin{align}\label{eq:Lzl+1}
L(z_{\ell+1}) &= \frac{L(z_{\ell+1}^-)}{2^R}=\frac{\text{e}^{F(z_{\ell+1}^{-} -z_{\ell})}L(z_{\ell})}{2^{R}}\nonumber\\
&= \frac{\text{e}^{F(z_{\ell+1}-z_0)}}{2^{R(\ell+2)}}L(z_0^-)
\end{align}
and the estimation error satisfies
\begin{equation*}\label{eq:qerr_zl+1}
|e_x(z_{\ell+1})|_\infty\le\frac{L(z_{\ell+1})}{2}.
\end{equation*} 
Combining this relation with the boundedness of $x(t)$ (by hypothesis), we have
\allowdisplaybreaks
\begin{align}\label{eq:estxzl}
|\overline x(z_{\ell+1})|_\infty&\le |x(z_{\ell+1})|_\infty +|e_x(z_{\ell+1})|_\infty\nonumber\\
&\le  W+\frac{L(z_{\ell+1})}{2}\nonumber\\
&= W+\frac{L(z_0^-)}{2}\frac{\text{e}^{F(z_{\ell+1}-z_0)}}{2^{R(\ell+2)}}\nonumber\\
&\le W+\phi_{\max}(z_0^-)\frac{\text{e}^{F((\ell+1)\Delta+\kappa+\eta \Delta)/\sigma}}{2^{R(\ell+2)}}\nonumber\\
&< \left(1+\frac{\text{e}^{F(\kappa+\eta \Delta)/\sigma}}{2^{R}}\right)W< 2W
\end{align}
where the third inequality comes from \eqref{eq:L<z0} and Lemma \ref{lem:zmbd}, the fourth from \eqref{eq:WO} and the condition \eqref{eq:lem2con}, and the last inequality again from $\frac{\text{e}^{F(\kappa+\eta \Delta)/\sigma}}{2^{R}}< 1$ implied by (17).

We have proved that $|\overline x(z_0)|_\infty\le 2 W$ in (\ref{eq:ovexz0}). Moreover, by the analysis between (\ref{eq:err_zl}) and (\ref{eq:estxzl}), we have shown that if the estimate exists and satisfies $|\overline x(z_{\ell})|_\infty\le 2 W$ for $\ell\in \Z_{\ge 0}$, then $|\overline x(t)|_\infty\le O$ for $t\in ]z_{\ell},z_{\ell+1}[$, $|\overline x(z_{\ell+1})|_\infty\le 2 W$ and \eqref{eq:Lzl+1}. By induction, we arrive at \emph{Claim b)} for the time interval $[z_{\ell},z_{\ell+1}[$ with $\ell\in \Zzero$.

Now, we have that the estimate error satisfies $|e_{x}(t)|_\infty\le L(t)/2$ for all $t\ge 0$. 
Then for $t\in [z_{\ell},z_{\ell+1}[$, we have
\begin{align}\label{eq:errexponen}
|e_x(t)|_\infty &\le \frac{L(t)}{2}=\frac{\text{e}^{F(t-z_0)}L(z_0^-)}{2^{R(\ell+1)+1}} \nonumber\\
	& \le \text{e}^{F(z_{\ell+1}-z_0)}\frac{L(z_0^-)}{2^{R(\ell+1)+1}}\nonumber\\
& \le W\text{e}^{\frac{F(\kappa+\eta\Delta)}{\sigma}}\left(\frac{\text{e}^{\frac{F\Delta}{\sigma}}}{2^{R}}\right)^{\ell+1}=\gamma\lambda^{\ell+1}
\end{align}
where the last inequality is due to Lemma \ref{lem:zmbd}, \eqref{eq:L<z0} and \eqref{eq:WO}. The reals $\gamma$ and $\lambda$ are given by	\eqref{eq:lem_gamlam}.	
Hence, we have established \eqref{eq:lemqerr>z0}.
\qedp

By this result, the estimation error may increase between two successful information transmissions in view of (\ref{eq:qerr_xt}). However, from \eqref{eq:lemqerr>z0}, we observe that as the number of successful transmissions increases, the estimate error eventually decays to zero. This implies that the state is asymptotically reconstructed by the decoder.

The bound on the estimation error in Proposition \ref{lem:qrangebitcon} is characterized by the number of successful transmissions $l$ for $l\ge 0$. To compare the asymptotic estimation result with that in \citet{depersis2004} for the case without DoS and to discuss the side-effects of DoS on the number of quantization bits, we reform the bounds on the norm of the estimation error \eqref{eq:lemqerr<z0} and \eqref{eq:lemqerr>z0} in Proposition \ref{lem:qrangebitcon} with time $t$ as the argument. This leads us to the following corollary.
\begin{corollary}
	Under the conditions of Proposition \ref{lem:qrangebitcon}, we have
	\begin{align}\label{eq:lemqerr>z0_cw}
	|e_x(t)|_\infty\le  c\text{e}^{-\omega t}
	\end{align} 
	where
	\begin{align}\label{eq:c_omega}
	c& = W\text{e}^{(2\ln (2)R-F\Delta/\sigma)(\kappa/\Delta+\eta)}\nonumber\\ 
	\omega &= \frac{\ln (2)\sigma R-F\Delta}{\Delta}.
	\end{align}
\end{corollary}
\emph{Proof.} By Proposition \ref{lem:qrangebitcon}, \eqref{eq:lemqerr>z0} holds for $\ell\ge 0$. Note that \eqref{eq:zmbound} implies
\begin{equation*}
\ell+1>\frac{(z_{\ell+1}-z_0)\sigma-\kappa-\eta  \Delta}{\Delta}.
\end{equation*}
Combine it with \eqref{eq:lemqerr>z0}, for $t\in [z_{\ell},z_{\ell+1}[$ we obtain
\begin{align}\label{eq:ext_lem2}
|e_x(t)|_\infty&<\gamma\lambda^{\frac{(z_{\ell+1}-z_0)\sigma-\kappa-\eta  \Delta}{\Delta}}\nonumber\\
& < \gamma\lambda^{\frac{(t-z_0)\sigma-\kappa-\eta  \Delta}{\Delta}}\nonumber\\
&= \gamma \text{e}^{\frac{-|\ln\lambda|}{T}((t-z_0)\sigma-\kappa-\eta \Delta)}\nonumber\\
&= \gamma \text{e}^{\frac{|\ln \lambda|(\kappa+\eta \Delta+z_0 \sigma)}{\Delta}} \text{e}^{-\frac{|\ln\lambda|\sigma}{\Delta}t}\nonumber\\
&\le \gamma \text{e}^{\frac{2|\ln \lambda|(\kappa+\eta \Delta)}{\Delta}} \text{e}^{-\frac{|\ln\lambda|\sigma}{\Delta}t} =  c\text{e}^{-\omega t}
\end{align} 
where the last inequality is due to \eqref{eq:z0bound} in Lemma \ref{lem:zmbd} and the last equality
follows by substituting $\Delta$ and $\lambda$ from (21). After substituting \eqref{eq:lem_gamlam} into $c$ and $\omega$, we get \eqref{eq:c_omega}.

For $t\in [0,z_0[$, by \eqref{eq:lemqerr<z0} and \eqref{eq:lem_gamlam}, we have 
\begin{align}
|e_x(t)|_\infty&\le W < \gamma \text{e}^{\frac{|\ln \lambda|(\kappa+\eta \Delta)}{\Delta}} \nonumber\\
&= \gamma \text{e}^{\frac{2|\ln \lambda|(\kappa+\eta \Delta)}{\Delta}} \text{e}^{-\frac{|\ln \lambda|(\kappa+\eta \Delta)}{\Delta}}\nonumber\\
&\le  c\text{e}^{-\frac{|\ln \lambda|(\kappa+\eta \Delta)}{\Delta}}\nonumber\\
&= c\text{e}^{-\frac{|\ln \lambda|\sigma}{\Delta}\frac{(\kappa+\eta \Delta)}{\sigma}}\nonumber\\
&\le  c\text{e}^{-\omega z_0} \le c\text{e}^{-\omega t} \nonumber
\end{align}
where the fourth inequality follows from \eqref{eq:z0bound}. This completes the proof. \qedp

We have a few remarks comparing the estimation error and the required quantization bits with those in \citet{depersis2004} as follows.
\begin{remark}
	 It is emphasized that our result on the data rate bound is a generalization of that in the literature. In particular, when no DoS attack is present in the system, the bound reduces to that of Lemma 1 of \citet{depersis2004}. Compared with the condition there, we observe that DoS can directly influence the number of quantization bits. Since $R$ is lower bounded as \eqref{eq:lem2con} and $\sigma<1$, we have that the number of bits needed for exponential state estimate is larger than that of the nominal case in \citet{depersis2004}. The result regarding $R$ is also intuitive in the sense that  when the communication suffers from DoS attacks, the transmission attempts are more likely to be interrupted. This implies that the state estimate uncertainty may expand more between two successful transmissions. Hence we need more bits to compensate the expansion than in the nominal situation in \citet{depersis2004}. Moreover, if the frequency and/or duration of DoS increases ($\tau_D$ and/or $T$ is smaller), $\sigma$ decreases and $R$ increases. Hence, to estimate the state under more severe DoS attacks, we need more bits. 
	 
	 Similar arguments on the relation between DoS and the required communication rate can be found in \citet{wakaiki2017quantized} and \citet{Shuai_Stabilizationun} in the case of linear systems; there, a more explicit relation to the minimum data rate result (e.g., \citet[Theorem 2]{Shuai_Stabilizationun}, \citet[Theorem 3.2]{wakaiki2017quantized}) can be established. Note that in \eqref{eq:lem2con}, $R$ depends linearly on $F$, which suggests that more unstable systems with larger $F$ requires more bits. This implication is in alignment with the linear case.
	 \qedp 
\end{remark}
\begin{remark}
	We should further highlight the subtle effect of the DoS on the number of bits through the Lipschitz constant $F$. The definition of $F$ relies on the specific state and estimate evolution sets $\mathcal{B}(W)$ and $\mathcal{B}(O)$. From \eqref{eq:WO}, these two sets both depend on the time of the first successful transmission. As $W\ge X$ and $O\ge 2W$ are always true, and the Lipschitz constant is monotonically increasing as the sizes of $\mathcal{B}(W)$ and $\mathcal{B}(O)$ increase, we have that the Lipschitz constant $F$ in the presence of DoS is always larger than that of the nominal case. For this reason, the quantization bits in the presence of DoS also increase. Moreover, by Lemma \ref{lem:zmbd}, for more severe DoS attacks, the first successful transmission can be postponed. Hence the prescribed evolution sets of the state and estimate will be larger, which leads to a larger $F$ and then requires more bits in the quantization. \qedp 
\end{remark}
\begin{remark}\label{rem:decayrate}
	Our last remark is more technical regarding the performance of the estimation error in  \eqref{eq:lemqerr>z0_cw} and \eqref{eq:c_omega} in comparison with that of Lemma 1 in \citet{depersis2004}. Assume that with the same number of bits $nR$, the norms of the state estimate errors for the transmission with and without DoS both can decay to zero. Then, there exist $\omega_1$ and $\omega_2$ satisfying $	|e_x(t)|_\infty\le  c_1\text{e}^{-\omega_1 t}$ for DoS-present case and $	|e_x(t)|_\infty\le  c_2\text{e}^{-\omega_2 t}$ for DoS-free scenario, respectively. Then by \eqref{eq:c_omega}, we have $\omega_1=(\ln (2)\sigma R-F\Delta)/\Delta$.
Moreover, for the latter, Lemma 1 in \citet{depersis2004} proves $\omega_2=(\ln(2)R-F\Delta)/\Delta$. Recall $0< \sigma<1$, which implies that the estimate error for the transmission with DoS may decay more slowly than that for the nominal situation.  
	On the other way around, if one wants to design a system such that $|e_x(t)|_\infty$ decays exponentially with a given rate $\omega$, then in the worst case, the number of bits needed for the nominal situation is $nR=\frac{n(F+\omega)\Delta}{\ln 2}$ by Lemma 1 in \citet{depersis2004}. While for the situation that the communication suffers from DoS, by \eqref{eq:c_omega} the number of bits needed to maintain $\omega$ decay rate is $nR=\max\left\{\frac{n(F+\omega)\Delta}{\ln (2)\sigma},\frac{nF(\kappa+\eta \Delta)}{\ln (2) \sigma}\right\}$. Since $\sigma<1$, to maintain the same decay rate of $|e_x(t)|_\infty$, we may need more bits for the state transmission with DoS attacks.
	\qedp
\end{remark}

\section{Asymptotic Stabilization under DoS}\label{sec:asymtoticsta}	
In the previous section, we have shown that the system state can be estimated asymptotically under a sufficiently large number of bits. This raises the possibility of stabilizing the system by the encoded state feedback. However, the validation of Proposition \ref{lem:qrangebitcon} depends on the assumption that the state always evolves within a set where $|x(t)|_\infty\le W$. Hence before applying the result in Proposition \ref{lem:qrangebitcon}, we need to first select a proper set in which we would like the closed-loop system to evolve; then, we estimate the number of quantization bits which ensures that the quantization error is a sufficiently small value at the time when the state of the system closely approaches the boundary of the selected set. This guarantees that the derivative of the Lyapunov function under the encoded state feedback will remain negative and not influenced by the quantization error. Thus the state will always evolve in the selected set.

With the encoded state feedback control $u(t)=k(\overline x(t))$, the corresponding closed-loop system can be written as follows:
\begin{align}\label{eq:inputsolution}
\dot x(t)&= f(x,k(x))+f(x,k(\overline x))-f(x,k(x))\nonumber\\
&= f(x,k(x))+g(x,\overline x)(x-\overline x) 
\end{align}
with $g(x,\overline x)$ given as in \citet{depersis2004}. 

Let
\begin{equation}\label{eq:l_def}
l:=\alpha_2(\phi_{\max}(\overline z_0))
\end{equation}
and define the level set of states
\begin{equation}
\Gamma_l:=\{x:V(x)\le l\}.
\end{equation}
The reals $W$ and $U$ introduced in Section \ref{sec:dynquant} and Proposition \ref{lem:qrangebitcon} are then given as follows:
\begin{align}\label{eq:WOU}
&W=\alpha_1^{-1}(l+\delta),\quad O=\alpha_1^{-1}\circ\alpha_2\left(2W\right)\nonumber\\
& U=\max\limits_{x:|x|_\infty\le O}|k(x)|_\infty
\end{align}
with $\delta$ being an arbitrary positive number. The Lipschitz constant $F$ is chosen as in Section \ref{sec:dynquant}. 

Set
\begin{equation}\label{eq:Mdef}
M:=\max\limits_{x\in \mathcal{B}(W),\overline x\in \mathcal{B}(O)} \left|\frac{\partial {V}}{\partial x}(g(x,\overline x))\right|_\infty.
\end{equation}
In the following lemma, we show that the system state satisfies  $|x(t)|_\infty\le W$ in a non-empty time interval.
\begin{lemma}\label{lm:evolset}
	If the number of the quantization bits is chosen as $B= nR $, with $R$ satisfying the condition \eqref{eq:lem2con}, then there exists a finite time $\theta:=z_0+\delta/(M\gamma)$ such that, for all $t\in[0,\theta)$, $x(t)\in \Gamma_{l+\delta}$ and
	\begin{equation}\label{eq:lm4dvass}
	\frac{\partial V}{\partial x}f(x(t),k(\overline x(t)))\le -\alpha(|x(t)|_\infty)+M|x(t)-\overline x(t)|_\infty.
	\end{equation}
\end{lemma}
\emph{Proof.} In view of the definition of $\phi_{\max} (t)$ in (\ref{eq:phit}), and the facts that
$|x(t)|_\infty\le \phi_{\max}(t)\le \phi_{\max}(z_0^-)$
and
\begin{equation*}
\mathcal{B}(\phi_{\max}(z_0^-))\subseteq \mathcal{B}(\phi_{\max}(\overline z_0))\subset \Gamma_{l} \subset \Gamma_{l+\delta} \subset \mathcal{B}(W),
\end{equation*}
we have $|x(t)|_{\infty}\le W$ for $t\in [0,z_0[$. This shows that $\theta$ exists and satisfies $\theta\ge z_0$.

By the proof of Proposition \ref{lem:qrangebitcon}, if $R$ satisfies \eqref{eq:lem2con} and there exists a time instant $ t'\ge z_0$ such that $|x(t)|_\infty\le W$ holds for $t\in [z_0, t']$, then the state estimate exists and satisfies $|\overline x(t)|_\infty\le O$ for all $t\in [0, t']$. Hence we consider the largest time interval $\mathcal{I}$ over which the state $x(t)$ satisfies
\begin{align}
|x(t)|_\infty\le W, \quad \forall t\in \mathcal{I}.\nonumber
\end{align}

\emph{Case a)} The interval $\mathcal{I}$ is finite. Then there exists a time instant $ \bar t\in [z_0,+\infty)$  such that $\mathcal{I}=[0,\bar
t]$. Let $z_{\ell}=\max\{z_k:z_k\le\bar t, k\in \Zzero\}$. 
By \eqref{eq:errexponen} and condition \eqref{eq:lem2con}, for $t\in [z_0,\overline t]$, one has
\begin{align}
|x(t)-\overline x(t)|_\infty\le \gamma\lambda^{\ell+1}<\gamma. \nonumber
\end{align}
Hence, we have
\begin{align}\label{eq:lm4Vdot}
&\frac{\partial V}{\partial x}f(x(t,k(\overline x(t))))\nonumber\\
&=\frac{\partial{V}}{\partial x}f(x(t),k(x(t)))+\frac{\partial{V}}{\partial x}g(x(t),\overline x(t))(x(t)-\overline x(t))\nonumber\\
&\le \frac{\partial{V}}{\partial x}f(x(t),k(x(t))) +M|x(t)-\overline x(t)|_\infty\nonumber\\
&< -\alpha(|x(t)|_\infty)+M\gamma
\end{align} 
for $t\in [z_0,\overline t]$, where the equality is obtained by  (\ref{eq:inputsolution}) and the first inequality follows from the definition of $M$. 

Due to the negativeness of $-\alpha(|x(t)|_\infty)$ in \eqref{eq:lm4Vdot}, we have
\begin{equation}\label{eq:V(t)<V(0)}
V(x(t))<V(x(z_0))+M\gamma t.
\end{equation}
It is easy to see $\overline t\ge z_0+\delta/(M\gamma)$; otherwise we would have
\begin{equation*}
V(x(\overline t))<l+\delta 
\end{equation*}
which implies $|x(\overline t)|_\infty< W$ and contradicts the definition of $\overline t$. By \eqref{eq:V(t)<V(0)}, we also have $x\in \Gamma_{l+\delta}$ for all $t\in [0,\overline t]$. The existence of $\theta $ then follows immediately. The inequality \eqref{eq:lm4dvass} comes from that \eqref{eq:lm4Vdot} holds for all $t\in [0,\overline t]$.

\emph{Case b)} The time interval $\mathcal I$ is infinite, namely $\mathcal I=[0,+\infty)$. Then, the inequality \eqref{eq:lm4Vdot} always holds, which implies that $\theta$ exists and \eqref{eq:lm4dvass} is valid.\qedp

The next result shows that the quantization error at time $t=\theta$ can be guaranteed to be  small enough as long as the number of quantization bits is sufficiently large. Let $K=\max\{\frac{\delta \sigma}{M\gamma \Delta}-\frac{\kappa}{\Delta}-\eta,1\}$.

\begin{lemma}
	For any $\eps>0$, there exists a number of quantization bits $B=nR$ with $R>0$ satisfying \eqref{eq:lem2con} and the following condition 
	\begin{equation}\label{eq:lm5bitno}
	R\ge \frac{F \Delta}{\sigma \ln 2}+\frac{\ln(\gamma/\eps)}{K\ln 2}
	\end{equation}
	such that for all $t\in [0,\theta]$ with $\theta>z_0$, \eqref{eq:lemqerr<z0} and \eqref{eq:lemqerr>z0} with $z_\ell=\max\{z_l:z_l\le \theta\}$ hold. And the quantization error at $t=\theta$ satisfies
	\begin{equation}\label{eq:qerrthetaeps}
	|e_x(\theta)|_\infty<\eps.
	\end{equation}
	Moreover, if $x(t)\in \Gamma_{l+\delta}$ for $t \in [\theta, \tilde \theta]$ with any $\tilde \theta >\theta$, 
	then 
	\begin{equation}\label{eq:qerrtidtheta}
	|e_x(t)|_\infty<\eps, \quad \forall t\in [\theta, \tilde \theta].
	\end{equation}
\end{lemma}
\emph{Proof.} Consider the time interval $\mathcal{I}$ introduced in the proof of Lemma \ref{lm:evolset}. We only focus on the situation that $\mathcal{I}=[0,\theta]$ with $z_0< \theta <+\infty$, since the result can be extended easily to the semi-infinite case. The first part that \eqref{eq:lemqerr<z0} and \eqref{eq:lemqerr>z0} hold comes directly from the fact that $R>\frac{FT}{\ln 2 \sigma}$ and Proposition \ref{lem:qrangebitcon}.

For the second part regarding \eqref{eq:qerrthetaeps}, by Lemma \ref{lm:evolset}, we know that $\theta$ exists and satisfies $\theta=z_0+\delta/(M\gamma )\in [z_{\ell},z_{\ell+1}[$ for some $\ell\in \Z_{\ge 0}$. Suppose the condition
\begin{equation}\label{eq:qerrthetabd<eps}
\gamma\lambda^{K}<\eps
\end{equation}
holds. Here we consider two cases depending on the value of $K$. First, if $\frac{\delta \sigma}{M\gamma \Delta}-\frac{\kappa}{\Delta}-\eta\le 1$, then $K=1$. Since $\theta>z_0$, $\ell\ge0$,  then by \eqref{eq:lemqerr>z0},
\begin{equation*}
|e_x(\theta)|_\infty \le \gamma\lambda^{\ell+1}\le \gamma\lambda^{K}\le \eps.
\end{equation*}
On the other hand, if $\frac{\delta \sigma}{M\gamma \Delta}-\frac{\kappa}{\Delta}-\eta>1$, then $K=\frac{\delta \sigma}{M\gamma \Delta}-\frac{\kappa}{\Delta}-\eta$. By the second inequality of \eqref{eq:ext_lem2}, we have
\begin{align}\label{eq:qerrtheta<eps}
|e_x(\theta)|_\infty&< \gamma\lambda^{\frac{(\theta-z_0)\sigma-\kappa-\eta  \Delta}{\Delta}}\nonumber\\
&= \gamma\lambda^{\frac{(z_0+\delta/(M\gamma)-z_0)\sigma-\kappa-\eta \Delta}{\Delta}}\nonumber\\
&= \gamma\lambda^{(\frac{\delta \sigma}{M\gamma \Delta}-\frac{\kappa}{\Delta}-\eta  )} 
=\gamma \lambda^{K}<\eps. 
\end{align}

Inequality \eqref{eq:qerrthetabd<eps} leads us to 
\begin{align}
\gamma\left(\text{e}^{\frac{F \Delta}{\sigma}}/2^{R}\right)^{K}<\eps
&\Leftrightarrow   K(F \Delta/\sigma-R\ln{2})<-\ln(\gamma/\eps)\nonumber\\
&\Leftrightarrow R>\frac{F \Delta}{\sigma\ln 2}+\frac{\ln(\gamma/\eps)}{K\ln 2} \nonumber.
\end{align}
This shows the part as in \eqref{eq:qerrthetaeps}.

For the statement regarding \eqref{eq:qerrtidtheta}, it is only possible that $\tilde \theta\le \overline t$ since $x(t)\in \Gamma_{l+\delta}$ implies that $|x(t)|_\infty\le W$, contradicting the definition of $\overline t$. If $\tilde \theta< z_{\ell+1}$, then $|x(t)|_\infty\le W$ for $t\in [z_{\ell},\tilde \theta]$, which by Proposition \ref{lem:qrangebitcon} implies that $|\overline x(t)|_\infty\le O$ for $t\in [z_{\ell},\tilde \theta]$. 

Suppose $\frac{\delta \sigma}{M\gamma \Delta}-\frac{\kappa}{\Delta}-\eta<1$, then $K=1$. Since $\tilde \theta \in [z_{\ell},z_{\ell+1}[$, then by \eqref{eq:lemqerr>z0},
\begin{equation}
|e_x(\tilde \theta)|_\infty \le \gamma\lambda^{\ell+1}\le \gamma\lambda^{K}\le \eps \nonumber.
\end{equation}
Now if $\frac{\delta \sigma}{M\gamma \Delta}-\frac{\kappa}{\Delta}-\eta> 1$, then $K=\frac{\delta \sigma}{M\gamma \Delta}-\frac{\kappa}{\Delta}-\eta$. By \eqref{eq:qerrtheta<eps} and the second inequality of \eqref{eq:ext_lem2},
\begin{align}
|e_x(\tilde \theta)|_\infty &<\gamma\lambda^{\frac{(\tilde \theta-z_0)\sigma-\kappa-\eta  \Delta}{\Delta}}\nonumber\\
&< \gamma\lambda^{\frac{(\theta-z_0)\sigma-\kappa-\eta  \Delta}{\Delta}} <\eps  \nonumber.
\end{align}
Hence we have if $z_{\ell}\le \tilde \theta< z_{\ell+1}$, then \eqref{eq:qerrtidtheta} holds true. The conclusion for the case that $\tilde \theta \ge z_{\ell+1}$ can be proved in an analogous way as for the case of $z_{\ell}\le \tilde \theta< z_{\ell+1}$. \qedp

As discussed earlier, intuitively we may expect that if the quantization error is sufficiently small, the negativeness of the Lypunov function derivative will not be influenced (the derivative of the Lyapunov function is semi-negative in the selected set that contains the origin), and hence the system will evolve in the set. The next lemma establishes a concrete result following this intuition. 
\begin{lemma}\label{lm:Nbdfinal}
	Given a positive $\rho<l+\delta$, if the number of quantization bits $B=nR$ is such that
\begin{align}
	R>& \max \left\{\frac{F \Delta}{\sigma \ln 2}, \frac{F(\kappa+\eta \Delta)}{\sigma \ln 2}\right\}\label{eq:Nfinalbound1} \\
	R\ge & \frac{F \Delta}{\sigma \ln 2}+\frac{\ln\left(M\gamma/(\alpha\circ\alpha_2^{-1}(\rho))\right)}{K\ln 2} \label{eq:Nfinalbound2}
	\end{align}
	then $x(t)\in \Gamma_{l+\delta}$, and \eqref{eq:lemqerr<z0} and \eqref{eq:lemqerr>z0} hold for all $t\ge 0$.
\end{lemma}
\emph{Proof.} Suppose \eqref{eq:lm5bitno} is satisfied, then inequality \eqref{eq:qerrthetaeps} holds. Combine it with \eqref{eq:lm4Vdot}, we have
\begin{equation*}
\frac{\partial V}{\partial x}f(x(\theta,k(\overline x(\theta))))
< -\alpha(|x(\theta)|_\infty)+M\eps.
\end{equation*}
Now suppose that at time $t=\theta$, the state belongs to the set
\begin{equation*}
\Gamma_{c+\delta}^{\rho}:=\{x\in \R^{n}:\rho\le V(x)\le l+\delta\}
\end{equation*}
with $\rho$ satisfying
\begin{equation}\label{eq:rhoeps}
-\alpha\circ \alpha_2^{-1}(\rho)+M\eps< 0.
\end{equation}
Then $\dot V(x(\theta))<0$, and thus the Lyapunov function decreases at $t=\theta$. Hence we know $\tilde \theta $ exists. Suppose that $\tilde \theta< +\infty$ is the largest time at which $x(t)\in \Gamma_{l+\delta}$ holds, which implies $x(\tilde \theta^+)\notin \Gamma_{l+\delta}$. However, from \eqref{eq:qerrtidtheta} and $x(\tilde \theta)\in \Gamma_{l+\delta}^{\rho}$, one has $\dot V(\tilde \theta)<0$ and $V(x(\tilde \theta^+))\le V(x(\tilde \theta))$. This contradicts the definition of $\tilde \theta$, which shows that $\tilde \theta=+\infty$. Then by Proposition \ref{lem:qrangebitcon}, \eqref{eq:lemqerr<z0} and \eqref{eq:lemqerr>z0} hold for all $t\ge 0$.

By \eqref{eq:rhoeps}, the condition \eqref{eq:Nfinalbound2} can be derived by replacing $\eps$ in \eqref{eq:lm5bitno} with $\alpha\circ \alpha_2^{-1}(\rho)/M$. \qedp


Lemmas \ref{lm:evolset}--\ref{lm:Nbdfinal} and Proposition \ref{lem:qrangebitcon} together show that the state $x(t)$ and the estimate $\overline x(t)$ can always evolve within the prescribed evolution sets by using a sufficiently large number of quantization bits, which proves the hypothesis mentioned before. By Proposition \ref{lem:qrangebitcon}, this indicates that the state estimate $\overline x(t)$ approaches the state $x(t)$ asymptotically and paves the way to show that the system can be stabilized by the encoded state feedback. We now arrive at the main result of this paper.

\begin{theorem}
	Consider the nonlinear process in \eqref{eq:sys} with control actions in (\ref{eq:control}) and (\ref{eq:dynxest}) under periodic transmission interval $\Delta$. Suppose that Assumptions \ref{asm:stability} and \ref{asm:statebd} hold and the DoS attacks characterized in Assumptions \ref{asp:DoSFreq} and \ref{asp:DoSDurat} satisfy {$\sigma=1-\frac{1}{T}-\frac{\Delta}{\tau_D} >0$}. Let $\bar z_0=(\kappa+\eta\Delta)/\sigma$ and $l=\alpha_2(\phi_{\max}(\overline z_0))$, with $\phi_{\max}$ given in \eqref{eq:phit}. For any $\delta>0$ and arbitrary $\rho<l+\delta$, if the number of quantization bits $B=nR$ satisfies \eqref{eq:Nfinalbound1} and \eqref{eq:Nfinalbound2}, the closed-loop system is stable.
\end{theorem}
\noindent\emph{Proof.} The nonlinear system \eqref{eq:sys} with the feedback control $u=k(\overline x)$ can be written as in \eqref{eq:inputsolution}:
\begin{equation}\label{eq:thmpertbsys}
\dot x(t)=f(x,k(x))+g(x,\overline x)(x-\overline x).
\end{equation}
Without the second term, this system is asymptotically stable. From Lemma \ref{lm:Nbdfinal} and Proposition \ref{lem:qrangebitcon}, we know that if the number of quantization bits is sufficiently large such that condition \eqref{eq:Nfinalbound1} and \eqref{eq:Nfinalbound2} hold, the second term converges asymptotically to zero and the state $x(t)$ with $x(0)\in \mathcal{B}(X)$ is well defined and bounded for all $t\ge 0$. Hence, by Theorem 10.3.1 in \citet{Isidori1999}, we have that the perturbed system \eqref{eq:thmpertbsys} is also asymptotically stable, which implies that the closed-loop system is stable.\qedp

\begin{remark}
		The lower bound on $R$ in \eqref{eq:Nfinalbound1} guarantees that the state estimate evolves within a bounded set and the estimation error converges to zero. The lower bound in \eqref{eq:Nfinalbound2} ensures the system state to evolve within a pre-selected bounded set, which  is necessary for the convergence of the estimate error. Since $\rho$ can be chosen arbitrarily close to $l+\delta$, intuitively this seems to imply that a large $\delta$ can reduce the lower bound in \eqref{eq:Nfinalbound2}. However, this lower bound also depends on $F$ and $M$, which are non-decreasing functions of the parameters $W$ and $O$. A large $\delta$ would make these two parameters large, which possibly leads to large $F,M,\gamma$, and therefore a large lower bound on $R$.  Hence, it is difficult to characterize a tight infimal of the quantization bits.  \qedp
\end{remark}
\begin{remark}
We now compare the obtained lower bounds in \eqref{eq:Nfinalbound1} and \eqref{eq:Nfinalbound2} with those known in the literature. It is clear that in our result, $R$ is always larger than $\frac{F \Delta}{\sigma\ln 2}$. This is at least $1/\sigma$ times larger than the one obtained in \citet{liberzon2005stabilization}, which assumed ISS of the system for encoding errors and did not consider DoS  attacks under the same value of $F\Delta$. Another related work \citet{depersis2004}
	considered the case with the assumption on state feedback stabilizability under no DoS attacks. That paper also derived two lowers bounds on $R$, and we have the counterparts in  \eqref{eq:Nfinalbound1} and \eqref{eq:Nfinalbound2}. With the same bound $X$ on the initial state, $W$ and $ O$ given in \eqref{eq:WOU} are always no less than those in \citet{depersis2004},  and hence $F$ in this paper is no smaller than that in \citet{depersis2004}. This shows that the lower bound on $R$ in \eqref{eq:Nfinalbound1} is always larger than the first lower bound on $R$ in \citet{depersis2004} ($\frac{F}{\Delta\ln 2}$), provided that the sampling period is the same. The second lower bound on $R$ in \citet{depersis2004} is $\frac{F\Delta}{\ln 2}+\frac{\ln(MX/(\alpha\circ\alpha_2^{-1}(\rho)))}{1/(2MX\Delta)}$. Let $\delta =1/2$ and assume that $\Delta, l$ and $M$ are the same as those in \citet{depersis2004}. If $\frac{\sigma}{M\gamma \Delta}-\frac{\kappa}{\Delta}-\eta\ge 1$, then $K<1/(2MX\Delta)$ since $\gamma >W\ge X$. The lower bound on $R$ in \eqref{eq:Nfinalbound2} is larger than the second lower bound on $R $ in \citet[Proposition 1]{depersis2004}. For the case $\frac{\sigma}{M\gamma \Delta}-\frac{\kappa}{\Delta}-\eta< 1$, it is difficult to make similar comparisons
	between the two bounds. Finally, in comparison with the linearization-based results in \citet{Kato2019}, our approach is capable of adjusting to an arbitrarily chosen bound $X$ on the initial state since our method relies on a globally stabilizing nonlinear control law. 
	\qedp
\end{remark}
	
\section{Numerical Simulations} \label{sec:examp}
In this section, we present a simulation example. We consider the system dynamics in \citet{depersis2004} as
\begin{equation}\label{eq:sys_examp}
\dot x = x^2-x^3 +u
\end{equation}
which is marginally stable without control. Select the controller $u=-\frac{5}{4}x$ and the candidate Lypunov function $V(x)=\frac{1}{2}x^2$. Then $\alpha_1(x)=\alpha_2(x)=\frac{1}{2}x^2$ and
\begin{equation}
\dot V(x)=x^3-x^4-\frac{5}{4}x^2\le -x^2=-2V(x).
\end{equation}
Hence the controller can stabilize the system and $\alpha(x)=2\alpha_1(x)=x^2$.

Let $x(0)=0.5$, $X=0.65$ and the sampling period $\Delta=0.1$ second. The parameters of the DoS signal are set as
\begin{equation}
\kappa =0.300, \quad \eta=1.300, \quad T=2.222, \quad\tau_D=0.714.
\end{equation} 
Hence $\sigma=1-\frac{1}{T}-\frac{\Delta}{\tau_D}=0.41$ and $\overline z_0=\frac{\kappa+\eta\Delta}{\sigma}=0.8985$. From \eqref{eq:phit}, \eqref{eq:l_def} and \eqref{eq:WOU}, to estimate $W,O$ and $U$, we need to find $\phi_{\max}(\overline z_0)$. This is obtained by simulating the uncontrolled system to $t=\bar z_0$ with different initial states $x(0)\in \mathcal{B}(0.65)$ and take the maximal absolute value of all the solutions. Simualtion shows that $\phi_{\max}(\bar z_0)=0.8$, hence $l=0.32$. Choosing $\delta=0.0001$ leads to $W\approx 0.8$, $O\approx 1.6$ and $U\approx 2$. Then by \eqref{eq:sys_examp}, one can verify that for any $x,\overline x$ and $u$ satisfying $|x|\le 0.8$, $|\overline x|\le 1.6$ and $|u|\le 2$,
\begin{align}
|x^2-x^3+u-\overline x^2+\overline x^3-u|_\infty  
\le  6.88 |x-\overline x| \nonumber
\end{align}
which shows that the Lipschitz constant $F=6.88$. In the simulation for the closed-loop system, we let $F=7$. Then by  \eqref{eq:lem_gamlam}, we have $\gamma=4311.1$. Now consider the constant $M$ in \eqref{eq:Mdef}. By \eqref{eq:inputsolution}, since \eqref{eq:sys_examp} is an affine function of $u$, $g(x, \overline x)=\frac{5}{4}$. Hence $M=\max_{|x|\le 1,|\overline x|\le 2}\left|\frac{5}{4}x\right|=1 $.

\begin{figure}[h]
	\centering 
	\subfigure[Process state $x$]{\label{fig:simu_state_2}
		\includegraphics[width=3.11in]{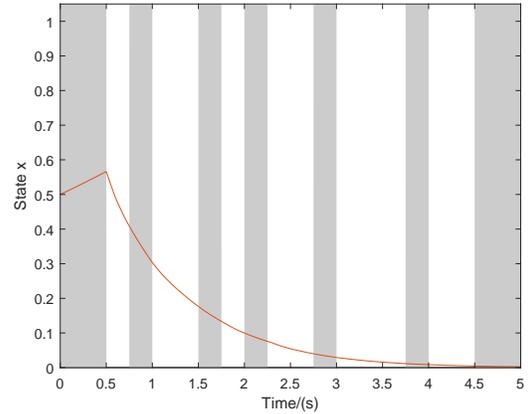}}
	\subfigure[Quantization range $L(t)$]{\label{fig:simu_J_2}
		\includegraphics[width=3.11in]{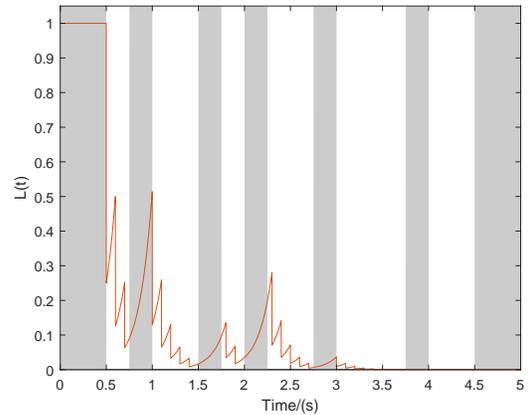}}
	\caption{System evolution with the quantized state-feedback control under DoS with $R=2$.}\label{fig:2bits}
\end{figure}

We are now ready to compute a lower bound on the quantization bit $R$, above which the state of the system can be stabilized to the origin. By \eqref{eq:Nfinalbound2}, if we select  $\rho=0.32$, a lower bound for $R$ can be computed and is equal to $16$. This bound is quite conservative, as in the simulation, if we select the quantization bits $R=2$, the closed-loop system is still stable for this example. The simulation result is shown in Fig. \ref{fig:2bits}, where the shaded areas represent the time intervals when the DoS attacks are present. From the first plot in Fig. \ref{fig:2bits}, we can see that after the first successful transmission, the state decreases to zero. By the second plot, one can see that $L(t)$ increases over the time intervals when the communication is not available and jumps to smaller values at the successful transmission instants. 



The conservativeness of the theoretical lower bound of the quantization bits may come from two reasons: a) the lower bound is derived based on the worst case where the inter-times of successful transmissions are always equal to the upper bounds implied from \eqref{eq:z0bound} and \eqref{eq:zmbound}, while this is not the case in the simulation; b) the estimate uncertainty does not enlarge as fast as that corresponding to the Lipschitz constant $F$.

%
%

\section{Conclusions}\label{sec:conclusion}
In this paper, we have designed a quantized state feedback controller to stabilize nonlinear systems, where the state transmissions are subject to DoS attacks. For state-feedback stabilizable nonlinear systems, we have shown that if the quantization bits are above a value which depends on the frequency and duration of the DoS signals, the systems can always be stabilized by the proposed quantized state feedback. As the bound in this paper is derived for general nonlinear systems, it may be conservative for certain classes of nonlinear systems, for example, systems whose dynamics are affine functions of the control inputs, as shown in the simulation.

For future works,  we would like to study reducing the quantization bits needed to stabilize nonlinear systems under DoS, possibly by adopting the ISS assumption in \citet{liberzon2005stabilization} or using the results in \citet{Claudio-nbit}. Moreover, one can consider systems with output measurements. Additionally, it is worthwhile to consider the situation of using dynamic controllers \citep{descusse1985decoupling,charlet1989dynamic}. However, the challenge may come from the design of a proper nonlinear dynamic controller that is easy to be adapted to incorporate the quantized discrete-time samples.

\bibliographystyle{myplainnast}
\bibliography{ifacconf}             



\end{document}